\documentstyle[sprocl,epsf]{article}

\def\sst{\scriptscriptstyle}

\def\be{\begin{equation}}
\def\ee{\end{equation}}
\def\bea{\begin{eqnarray}}
\def\eea{\end{eqnarray}}

\begin{document}

\title{SELF-AVOIDING POLYMERIZED MEMBRANES}

\author{ Fran\c cois DAVID }

\address{Service de Physique Th\'eorique, CE de Saclay\\
91191, Gif-sur-Yvette CEDEX, FRANCE}


\rightline{Saclay T95/111}
\rightline{cond-mat/9509096}
\vskip 3.ex
\maketitle\abstracts{
Recent progresses in the understanding of the scaling behavior
of self-avoiding flexible polymerized membranes (tethered manifolds) are  
reviewed (joint works with B. Duplantier and E. Guitter, and with K. Wiese).
They rely on a new general renormalization group approach for a
class of models with non-local singular interactions.
This approach allows to prove the existence of a $\epsilon$-expansion
for the scaling exponents, and validates the one loop results obtained
by direct renormalization methods.
Applications of the method to polymerized membranes at the tricritical
$\Theta$-point are presented.
}

\footnotetext[0]{STATPHYS19, Xiamen, China, 31 July - 4 August, 1995}
\section{What are polymerized membranes ?}
Polymerized membranes, also called tethered membranes, are an interesting
model for two-dimensional flexible polymerized films \cite{r:Jerus}.
In particular, they generalize the well-known problem of
the statistical behavior of long chains, i.e. of polymers.
However, even simple properties, such as the statistical properties at thermal
equilibrium of a single membrane, are quite difficult to study theoretically,
and require the development of new techniques.
I shall review some recent progresses made with B. Duplantier and E. Guitter
\cite{r:DDG3},
and some applications of these techniques with K. Wiese \cite{r:WD95}. 

Let me consider an ideal flexible two dimensional network fluctuating in three
dimensional space, characterized by:
{\it (i)} a fixed and regular internal structure
with a fixed connectivity (for instance a triangular lattice), and {\it (ii)} 
negligible bending rigidity.
The only forces between elements of the network are the elastic forces
(described for instance by a harmonic potential between nearest neighbors).
If self-avoiding interactions are not taken into account, one deals with a
``phantom" membrane, which is known to be, for vanishing or small enough bending
rigidity, in a crumpled phase.
The large scale fluctuations are described by a Gaussian model.
Such a model is physically different from the random surfaces used in  
models of fluid two-dimensional films (for instance for vesicles or
micro-emulsions).
For these models, the bending rigidity is important and the fluid character
of the film is reproduced by a fluctuating connectivity of the lattice.

Taking into account the self-avoiding interactions (which are short ranged in
the three dimensional embedding space, but involve elements of the lattice
which can be arbitrarily far apart in term of the internal lattice distance)
one expects that the long range properties of the membrane will be
changed.
For instance, the average size $\langle R\rangle$ of the membrane in
physical space is expected to scale with the internal size $L$ of the
lattice as
\begin{equation}
\langle R\rangle\ \propto\ L^\nu\qquad 0<\nu<1
\label{eNu}
\end{equation}
if the membrane is still crumpled, but swollen by self-avoidance.
The membrane may even become flat (this is observed in numerical
simulations of self-avoiding tethered membranes in three dimensions);
the exponent $\nu$ is then equal to $1$.

\section{An Edwards Model for $D$-dimensional polymerized membranes:}
In the continuous model introduced in \cite{r:AroLub87,r:KarNel87},
the field $\vec r(x)$ describes the embedding of a $D$-dimensional space,
whose points are labeled by the coordinate $x$,
into the external $d$-dimensional space, whose points are labeled by the vector
$\vec r$.
The Hamitonian $H$ (the free energy for a configuration $r$) is the sum of a
Gaussian
elastic energy and of a 2-body repulsive interaction, proportional to the
coupling constant $b$:
\begin{equation}
\label{eHam}
H[\vec r]\ =\ \int d^D\! x\  {{1\over 2}} (\nabla_{\!x}\vec r) ^2\ +\
b\ \int d^D\!x\int d^D\!y\ \delta^{d}(\vec r(x)-\vec r(y))
\ .
\end{equation}
The internal dimension $D$ may be taken as a continuous parameter,
interpolating between polymers ($D=1$) and membranes ($D=2$).
The issue is to compute with this model critical exponents describing the
scaling behavior of large membranes, for instance the exponent $\nu$
(related to the fractal dimension $d_f$ of the membrane by $\nu=D/d_f$), and the
configuration exponent $\gamma$, related to the scaling of the partition
function $Z$ of a finite membrane with internal extent $L$ by
\begin{equation}
Z\ \propto\ L^{\gamma-1}\ \hbox{\rm constant}^{L^D}\ .
\label{eGamma}
\end{equation}
The mean field exponents are obtained by setting $b=0$ and one obtains
$\nu_{0}=(2-D)/2$ and $\gamma_{0}=1-d(2-D)/2$.

Dimensional analysis shows that mean field theory is invalid if $\epsilon$,
the engineering dimension of $b$, is positive
\begin{equation}
[b]\ =\ \epsilon\ =\ 2D-d(2-D)/2\ >\ 0\ .
\end{equation}
\begin{figure}
\centerline{\epsfbox{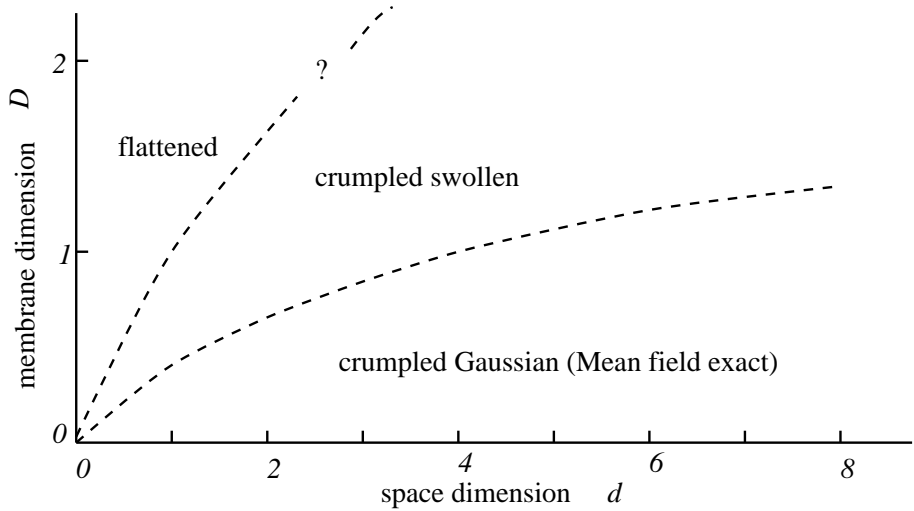}}
\caption{$d$-$D$ plane}
\label{fdDplane}
\end{figure}
The general picture of the expected behavior as a function of the internal
dimension $D$ and of the external dimension $d$ is presented in Fig.
\ref{fdDplane}

A natural idea is to compute the corrections to mean field by a
$\epsilon$-expansion \`a la Wilson-Fisher.
Using the so-called direct renormalization, inspired from polymer
theory \cite{r:Clo81}, the first
explicit calculations have been performed to first order in $\epsilon$
by Aronowitz \& Lubensky \cite{r:AroLub87}\ 
and by Kardar \& Nelson \cite{r:KarNel87} .
The basic idea of this method is to perform explicit perturbative calculations
for a finite membrane and for $\epsilon>0$.
Perturbation theory is then UV and IR finite, but has UV divergences when
$\epsilon\to 0$.
These poles in $1/\epsilon$ can be removed by reexpressing the observables in
terms of adequate dimensionless renormalized quantities, such as the 2nd virial
coefficient.
The internal size $L$ of the membrane plays the role of a renormalization ${\rm
mass}^{-1}$ scale, and the renormalization group equations can thus be obtained,
by writing the $L$ dependence of the renormalized theory.
The consistency of this procedure at first order has been checked by
explicit calculations by Duplantier, Hwa \& Kardar \cite{r:Dup&al90},
but its consistency at all
orders cannot be proved by the de Gennes trick valid for polymers
(the scaling limit of the self-avoiding walk can be mapped into a local field
theory with O($n$) symmetry in external $d$-dimensional space, in the limit
$n\to 0$, and standard renormalization group theory is then appliquable).

\section{Renormalization Theory and Renormalization Group for non-local
Theories:}
Instead of using direct renormalization, we succeeded in proving the
renormalizability of the model (\ref{eHam}), by considering it
as a non-local field
theory in $D$ dimensions \cite{r:DDG3}.
Perturbation theory is obtained by expanding the observables as power series
in $b$.
The bi-local ``interaction vertex" (in field theoretic language I denote it a
bilocal operator) is written in Fourier transform as
\begin{equation}
\raisebox{-3.ex}{\epsfbox{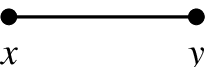}}\ \ =\ \delta^{d}(\vec r(x)-\vec r(y))\ =\ \int d^d\!k\ 
{\rm e}^{i\vec k(\vec r(x)-\vec r(y))}\ .
\label{eBilOp}
\end{equation}
It can be viewed in a Coulomb gas representation as the integral over the
``charge" $\vec k$ of a neutral ``dipole" with charge $+\vec k$ at $x$ and
charge $-\vec k$ at $y$.
The term of order $b^K$ in the perturbative expansion of the partition function
(as well as of other observables) involves $K$ dipoles
$(x_1,y_1),\ldots\,(x_K,y_K)$.
The integration over the charges $\vec k_1,\ldots\,\vec k_K$ gives an integral
over the positions of the dipoles of the determinant of the ``dipole energy"
quadratic form $Q$
\begin{equation}
\int\cdots\int \prod_{i=1}^{K}\ d^D\!x_i\,d^D\!y_i\
\det\Big[Q[x_i,y_i]\Big]^{-d/2}\ .
\end{equation}
$Q$ is a $K\times K$ matrix such that $\sum\limits_{i,j=1}^{K}\vec k_i
Q_{ij}\vec k_j$ is the Coulomb energy (in $D$-dimensions) of the $K$ dipoles.
Each $Q_{ij}$ is a linear combination 
\begin{equation}
Q_{ij}\ =\ G_0(x_i,x_j)+G_0(y_i,y_j)-G_0(x_i,y_j)-G_0(x_j,y_i)\ .
\label{eGzero}
\end{equation}
of the Coulomb potentials $G_0$ between endpoints of the dipoles $i$ and $j$
\begin{equation}
G_0(x,x')\ =\ \langle r(x)r(x')\rangle_{\!0}\ =
 {\Gamma\big((D-2)/2\big)\over 4\pi^{D/2}}\  |x-x'|^{2-D}
\ \ .
\label{eCoulPot} 
\end{equation}
The Coulomb potential is properly defined for $0<D<2$ by analytic continuation.
It is then negative, it vanishes for $x=x'$ (for $D>2$ it diverges), but as for
$D>2$ is decreases at large distances.
The integration over the $2K$ points in non-integer $D$-dimensional space can
also be defined properly by analytic continuation in $D$ and the use of distance
geometry.
This amounts to replace the integration over the $2K\times D$
coordinates by an integration over the $K\times (2K-1)$ scalar
distances between these points.

One can show that short distance UV singularities occur in the integrals, when
the determinant $\det[Q]$ vanishes.
This occurs if and only if some of the end-points of (not necessarily the same)
dipoles coincide, forming ``atoms", and so that the dipoles form ``molecules,
and moreover if one can assign non-zero charges $\vec k_i$ to the dipoles 
while the atoms stay neutral. 
This condition is more easily depicted graphically on Figure 2.
\begin{figure}
\centerline{\epsfbox{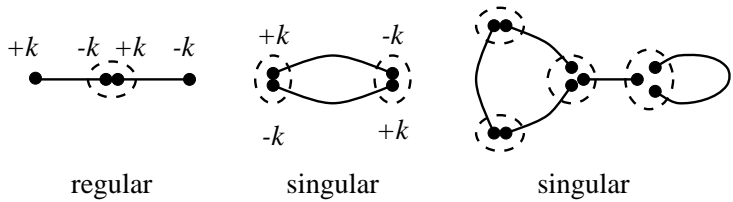}}
\caption{A non-neutral regular configuration and two neutral UV singular
configurations}
\end{figure}

The associated singularities of these integrals are related to the
behavior at short distance of the expectation value (with respects to the free
Gaussian model) of products of bilocal operators as given by Equ.~\ref{eBilOp}.
One can show that this short distance behavior is encoded in a multilocal
product expansion (MOPE), which generalizes Wilson's operator product
expansion.
Let me give two examples:

When the two points $x$ and $y$ of the bi-local interaction operator tend
towards a single point, this operator can be expanded in terms of local
operators involving derivatives of the field $\vec r$.
The first terms of the expansion are explicitly (omitting unimportant
coefficients)
\begin{eqnarray}
&&\vbox{\vskip 30pt}\nonumber\\
\ \ &=\ &|x-y|^{\epsilon-2D}\ {\bf 1}\nonumber \\
\raisebox{-0.ex}{\epsfbox[0 0 27 0]{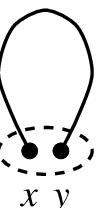}}
&& +\ |x-y|^{\epsilon-D-2}(x^\alpha-y^\alpha)(x^\beta-y^\beta)\
:\nabla_{\!\alpha}\vec r\,\nabla_{\!\beta}\vec r:\nonumber \\
&&+\ \cdots
\label{eMOPE1}
\end{eqnarray}
${\bf 1}$ is the identity operator (its expectation value is 1),
the $:\ :$ in the operator $:\nabla_{\!\alpha}\vec r\,\nabla_{\!\beta}\vec r:$
denotes the normal ordering subtraction prescription required to deal properly
with the UV singularities contained in $\nabla_{\!\alpha}\vec
r\,\nabla_{\!\beta}\vec r$.

The second example is less simple, and shows that when the end-points of two
bilocal operators tend pairwise towards two different points, this generates
again bilocal operators
\begin{eqnarray}
&=\ &\left[|x_1-x_2|^{2-D}+|y_1-y_2|^{2-D}\right]^{-d/2}\ 
\epsfbox{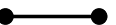}
\nonumber \\
\raisebox{-0.ex}{\epsfbox[0 0 92 0]{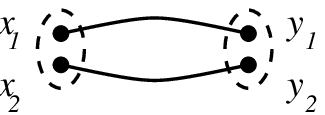}}\ 
&&\ +\ \cdots
\end{eqnarray}

This structure is generic, and products of local and bilocal operators generate
multilocal operators of the general form
\begin{equation}
\Phi\{x_1,\cdots,x_P\}\ =\ \int d^d\!\vec r_0\ \prod_{i=1}^P\
\left[\left(\nabla_{\vec r_0}\right)^{m_i}\ \delta^d(\vec r_0-\vec r(x_i))\
A_i(x_i)\right]\ .
\label{eMulOp} 
\end{equation}
where the $A_i(x_i)$ are local operators, which can be decomposed into
products of multiple $x$-derivatives of $\vec r$.
The $m_i$ are integers.
For $P=1$ and $m=0$ one recovers local operators $A(x)$ ($m>0$ gives $0$).
For $P=2$, $m_1=m_2=0$ and $A_1=A_2={\bf 1}$ one recovers the bilocal
interaction operator, etc$\ldots$ 
These operators have a very special form: they can be viewed as a local
convolution in the external $d$-dimensional  $\vec r$ space of a non-local
product (in the internal $D$-dimensional space) of the $P$ local operators
$A_i$.

The MOPE implies that the formalism of renormalization theory and of
renormalization group equations, which has been developed for local quantum
field theories, can be adapted for this model.
One is in fact interested in the IR scaling behavior of the lattice model,
when some length scale $L$ goes to $\infty$.
This lattice model is described by the Hamiltonian (\ref{eHam}), with a
short distance lattice cut-uff $a$.
To study this IR limit it is equivalent to look at the UV continuum limit of
the model when the physical length scale $L$ is kept fixed, while the UV
cut-off $a$ goes to $0$.
In this limit one can construct, via renormalization, a finite renormalized
theory with $a=0$, which obeys renormalization group equations.
From these equations, one recovers the large distance behavior of the
lattice model I started from.
The procedure works well in perturbation theory when one is close to the upper
critical dimension, i.e. for $\epsilon$ small, and it
leads to the $\epsilon$-expansion.

\begin{figure}
\centerline{
 \hbox{
  \epsfbox{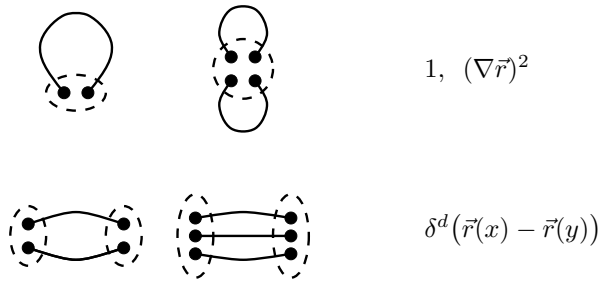}\kern10.ex
  \vbox{
   \hbox{1,\ \ $(\nabla\vec r)^2$}
   \hbox{\vspace{8.5ex}}
   \hbox{$\delta^d\big(\vec r(x)-\vec r(y)\big)$}
   \hbox{\vspace{1.ex}}
  }
 }
}
\caption{UV divergent configurations and the associated relevant operators}
\label{fDiv}
\end{figure}
In our case, the MOPE can be used to determine, by power counting, which
multilocal operators are relevant and give UV singularities (poles in
$1/\epsilon$).
Then one can also show that these poles can be subtracted by adding to the
Hamiltonian (\ref{eHam}) counterterms proportional to the marginally
relevant multilocal operators, leading to the UV finite renormalized theory.
For the model of self-avoiding surfaces, this analysis shows that the
UV divergences
are associated only with local and bilocal operators,  as depicted on
Fig.~\ref{fDiv}, and that only three
operators are relevant:
the identity operator ${\bf 1}$, the elastic energy operator ($\nabla\vec r)^2$
and the bilocal operator $\delta^d(\vec r(x)-\vec r(y))$.
${\bf 1}$ is strongly relevant, and gives power-like UV divergences
proportional to $a^{-D}$ ($a$ being a short-distance cut-off).
The two other operators are superficially relevant, they give logarithmic
UV divergences or equivalently poles in $1/\epsilon$ at $\epsilon=0$.

\begin{figure}
\centerline{\epsfbox{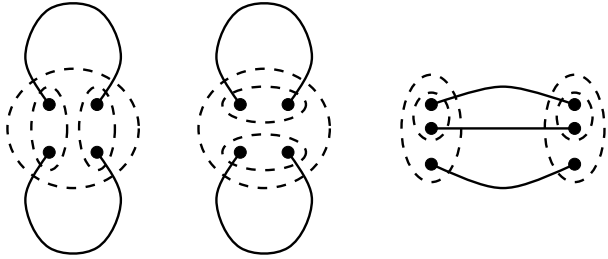}}
\label{fNesDiv}
\caption{Examples of nested singular configurations}
\end{figure}

The fact that the so-called superficial divergences, associated to a global
contraction of points towards a singular configuration, can be subtracted by
counterterms is a  consequence of the MOPE.
A complete proof of the renormalizability of the theory is possible, but much
more delicate.
It requires a control of the subdivergences coming from successive contractions
associated to nested singular configurations, such as those depicted of
Fig.~\ref{fNesDiv} .

\section{Scaling for infinite self-avoiding membranes}
A first application of this formalism is the derivation of scaling laws.
Since the model is renormalizable (at least perturbatively),
it can be made UV finite (for $\epsilon\simeq 0$) by introducing two
counterterms in the Hamiltonian.
The new renormalized Hamiltonian is of the form
\begin{equation}
H[\vec r]\ =\ {Z\over 2}\,\int d^D\!x\ (\nabla \vec r)^2\ +\ b_R\,\mu^\epsilon\ 
Z_b\ \int\int d^D\!x\,d^D\!y\ \delta\big(\vec r(x)-\vec r(y)\big)\ .
\label{eRenHam}
\end{equation}
$b_R$ is the dimensionless renormalized coupling constant (the perturbative
expansion in $b_R$ is UV finite order by order).
$Z$ is a ``wave-function" renormalization factor and $Z_b$ a coupling constant
renormalization factor, both are perturbative series in $b_R$, with poles up to
degree $1/\epsilon^{K-1}$ at order $K$.
$\mu$ is the renormalization momentum scale.
As for ordinary local theories, such as the Landau-Ginzburg-Wilson $\Phi^4$
Hamiltonian, one can change $b_R$ and $\vec r$ in Equ.~\ref{eRenHam} into bare
quantities in order to rewrite the renormalized Hamitonian as a bare Hamiltonian
given by Equ.~\ref{eHam}.
The renormalization group $\beta$-function and the anomalous dimension $\gamma$
of the field $\vec r$ are now defined in the standard way
\begin{equation}
\beta(b_R)\ =\
\left.\mu{\partial\over\partial\mu}b_R\right |_{\sst bare}
\qquad;\qquad
\gamma(b_R)\ =\ -\,{1\over 2}
\left.\mu{\partial\over\partial\mu}\ln
Z\right |_{\sst bare}\ .
\end{equation}
\begin{figure}
\centerline{\epsfbox{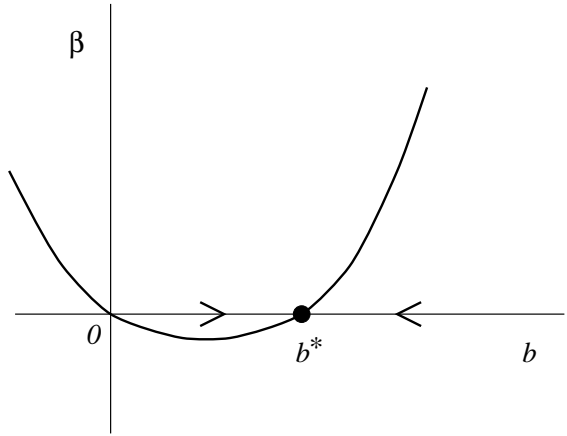}}
\label{fBeta}
\caption{The $\beta$-function and the RG flow for $\epsilon>0$.}
\end{figure}

The $\beta$-function is found to be of the form
\begin{equation}
\beta(b_R)\ =\ -\epsilon\, b_R+\hbox{\bf c}\,b_R+{\cal O}(b_R^2)
\qquad;\qquad\hbox{{\bf c} positive constant}\ ,
\label{eBeta}
\end{equation}
and therefore there is, at least for small $\epsilon>0$, an IR attractive 
fixed point $b_r^\star={\cal O}(\epsilon)$, which governs the scaling behavior
of self-avoiding polymerized surfaces at large distance.
The existence of this fixed point ensures the universality of this
non-trivial scaling for $\epsilon>0$i, and that
no new interactions, possibly non-local in
external space, are generated by the RG transformations.

The explicit calculation
leads to the same results for the scaling exponents $\nu$ and
$\gamma$ than the direct renormalization method at first order in $\epsilon$.
With this method higher order calculations are in principle feasible, but
technically quite difficult.
In particular, already at second order the RG functions cannot be expressed
analytically, and numerical integration methods have to be developed.
Work is in progress to compute the scaling exponents at order $\epsilon^2$.

\section{Finite Size Scaling and Direct Renormalization}
The model given by Equ.~\ref{eHam} describes an infinite membrane with flat
internal metric, corresponding to an infinite and regular flexible lattice.
Finite membranes are described by a similar model, but the $D$-dimensional
membrane $M$ is now embodied with a fixed non-trivial Riemannian metric 
$g_{\alpha\beta}(x)$, with curvature $R$ (examples are closed membranes with the
topology of the sphere ${\cal S}_D$ or the torus ${\cal T}^D$), and may have a
boundary $\partial M$ (open membrane with the topology of the disk for
instance).
A similar analysis can be performed for such models, and the MOPE structure of
short distance singularities is still valid, but new local operators $A(x)$,
which depend on the internal metric on $M$ and on the boundary $\partial M$,
such as the curvature $R$, appear in the MOPE and in Equ.~(\ref{eMulOp}).
The renormalized Hamiltonian now contains at least five operators and five
independent renormalization factors $Z$
\begin{eqnarray}
H[\vec r]\ =\ &&\int_M\,Z\,\hbox{\bf 1}\ +\ \int_M\,Z\,(\nabla\vec
r)^2\ +\ \int\!\int_M\,Z\,b\,\delta^d(\vec r-\vec r)\nonumber\\
&&+\ \int_M\,Z\,R\ +\ \int_{\partial M}Z\,\hbox{\bf 1}\ \ .
\label{fRenHam2}
\end{eqnarray}
The curvature operator $\int_M R$ is superficially relevant only for $D=2$ and
the boundary operator $\int_{\partial M}\hbox{\bf 1}$ only for $D=1$.
When these additional terms are not relevant, the first three renormalization
factors $Z$ are the same for finite curved membranes than for the infinite flat
membrane.
This property is analogous to the renormalization property of local field
theories in finite geometries, which justifies the finite scaling laws for
critical systems in finite geometries, and it has two very important
consequences:
{\it (i)} The scaling hypothesis at the basis of the direct renormalization
approach, which relies explicitly on calculations with finite membranes, is
shown to be valid to all orders in perturbation theory;
{\it (ii)} for ''abstract" membranes with dimension $D<2$, with the only
exception of open polymers ($D=1$ open surface), the following hyperscaling relation
\cite{r:Dup87}\ relating the configuration and the $\nu$ exponents holds:
\begin{equation}
\gamma\ =\ 1\,-\,\nu\,d\ \ .
\label{eHypSc}
\end{equation}

\section{Self-avoiding Membrane at the tricritical $\Theta$-point:}
\begin{figure}[t]
\centerline{\epsfbox{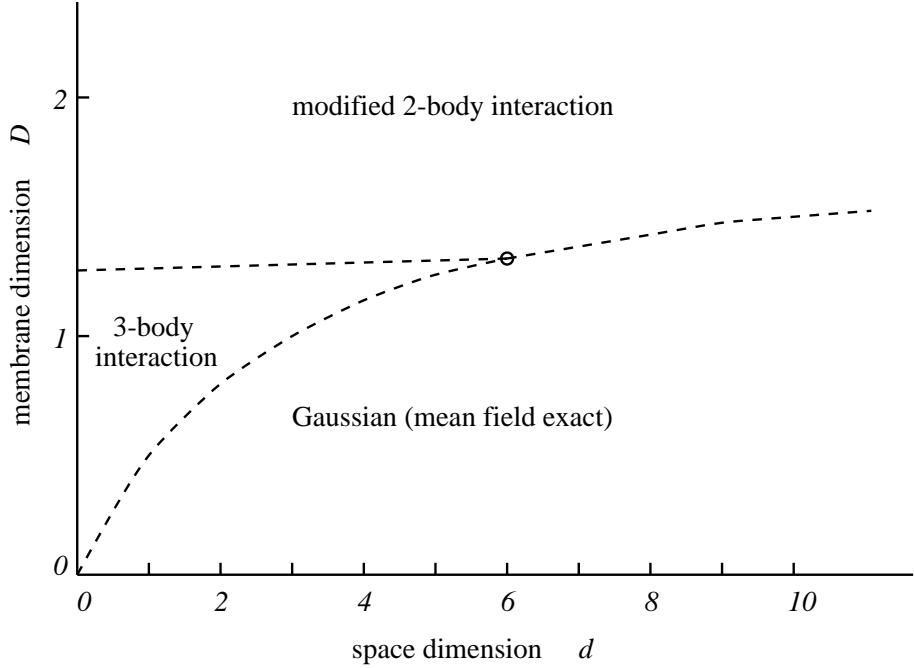}}
\label{fdDTheta}
\caption{Relevant interactions for the $\Theta$-point in the $d$-$D$ plane.}
\end{figure}
Finally, let me briefly discuss recent results obtained with K. Wiese on the
scaling behavior of polymerized membranes at the $\Theta$-point \cite{r:WD95}.
This point separates the swollen phase,
where the self-avoidance repulsive forces that I considered previously
dominate, from the dense collapsed phase, where short ranged attractive forces
dominates.
At the $\Theta$-point the effective two body repulsive coupling $b$ vanishes,
and two different interactions may become relevant.
The first one is the 3-body contact repulsion, which is usually considered for
polymers
\begin{equation}
\raisebox{-6.ex}{\epsfbox{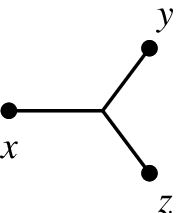}}\  
\ =\ \int\!\int\!\int\,d^D\!x\ d^D\!y\ d^D\!z\ 
\delta^d\big(\vec r(x)-\vec r(y)\big)\,\delta^d\big(\vec r(x)-\vec r(z)\big)
\ .
\label{f3Int} 
\end{equation}
The second one is a modified 2-body interaction, repulsive at short range but
attractive at larger range ($\Delta_{\vec r}$ is the $d$-dimensional Laplacian)
\begin{equation}
\raisebox{-2.ex}{\epsfbox{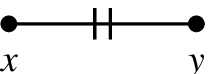}}\  
\ =\ -\ \int\!\int\,d^D\!x\,d^D\!y\ \Delta_{\vec r} \,
\delta^d\big(\vec r(x)-\vec r(y)\big)\ .
\label{f2'Int}
\end{equation}
Calculations at first order are not feasible analytically, and already require
numerical evaluations of complicated integrals.
The results of such one loop calculations are schematically depicted on
Fig.~\ref{fdDTheta}, where the domains where the 3-body and modified
2-body terms are respectively relevant are shown.
This indicates that the last modified 2-body term is the relevant one for
2-dimensional membranes in any external dimension $d$.
There is also a quite interesting and non-trivial crossover between the two
terms around $D=4/3$ $d=6$, which must be studied by a double
$\epsilon$-expansion.

\section{Conclusion:}
The theoretical study of the scaling behavior of polymerized flexible membranes
leads to the development of new multilocal continuum field theories, and to new
applications of renormalization group methods.
I hope that these methods will lead to a quantitative progress in the
understanding of the behavior of real 2-dimensional polymerized membranes.
This requires results beyond first order in the
$\epsilon$-expansion (recall that $D=2$ correspond to $\epsilon=4$), and
a better understanding of the relation between this RG approach and more
heuristic or approximate methods, such as variational methods or approximate
recursion relations.
The sophisticated renormalization theory for multilocal models should
hopefully also find  applications in other problems of statistical physics,
or in other areas of theoretical physics.

\end{document}